\documentclass[aip,apl,graphicx]{revtex4-1}
\usepackage{epsfig}
\draft 

\begin{document}


\title{On-Chip Josephson Junction Microwave Switch} 



\author{O. Naaman}
\email[]{ofer.naaman@ngc.com}
\author{M. O. Abutaleb}
\author{C. Kirby}
\author{M. Rennie}
\affiliation{Northrop Grumman Systems Corp., Baltimore, Maryland 21240, USA }


\date{\today}

\begin{abstract}
The authors report on the design and measurement of a reflective single-pole single-throw microwave switch with no internal power dissipation, based on a superconducting circuit containing a single Josephson junction. The data demonstrate the switch operation with 2 GHz instantaneous bandwidth centered at 10 GHz, low insertion loss, and better than 20 dB on/off ratio. The switch's measured performance agrees well with simulations for input powers up to -100 dBm. An extension of the demonstrated circuit to implement a single-pole double-throw switch is shown in simulation. 
\end{abstract}


\maketitle 


Quantum integrated circuits based on solid state qubits with increasing complexity have been demonstrated recently \cite{Kelly15,Corcoles15}, driving a growing need for microwave control technologies that can operate in the cryogenic environment with little or no dissipation and minimal losses in the signal path. A complete toolbox of integrated  microwave components must include amplifiers, circulators, mixers, and switches. Amplifiers \cite{Abdo14,OBrien14,Mutus13} and circulators \cite{Silwa15,Kerckhoff15} have already been realized in Josephson junction based superconducting circuits; cryogenic semiconductor microwave switches have been demonstrated \cite{Hornibrook15}, and switches based on parametrically pumped Josephson devices have been proposed \cite{Silwa15,Kerckhoff15}. In this Letter, we report on Josephson junction based microwave switches that represent a family of integrated devices for routing and modulation of microwaves. These devices have low insertion loss, over 20\% instantaneous bandwidth, and no internal power dissipation. In contrast with the switching solutions proposed in Refs. \onlinecite{Silwa15,Kerckhoff15,Hornibrook15}, our devices require no additional pump tones, and are driven by $\Phi_0$-level base-band flux signals, making them compatible with single flux quantum (SFQ) digital controllers \cite{Herr11,Mukhanov11}. 

\begin{figure}
\includegraphics[width=3.2in]{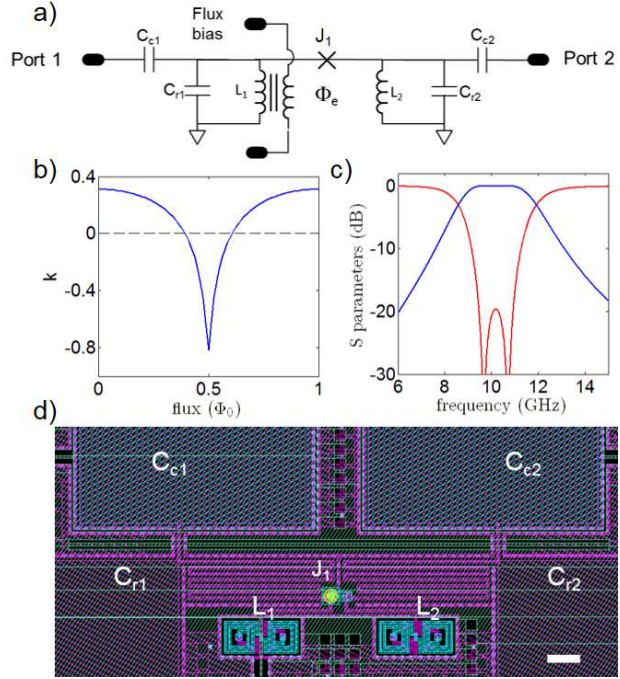}
\caption{\label{fig1}a) Circuit schematic of the Josephson junction microwave switch, b) effective mutual inductance of an RF-SQUID coupler with $\beta_L=0.9$, c) calculated transmission S$_{21}$ (blue) and reflection S$_{11}$ (red) of the device at $\Phi=0$, and d) CAD drawing of the device. Scale bar is $5~\mu$m.}
\end{figure}

Our implementation of a single-pole single-throw (SPST) switch, arguably the most basic device for integrated routing and modulation of microwave signals, is shown schematically in Figure \ref{fig1}a. At the heart of this device is an rf-SQUID variable coupler of the type described in Ref.\ \onlinecite{Chen14}, formed by the junction and inductors $L_1$ and $L_2$, which we take to be equal $L_1=L_2=L$. The rf-SQUID functions as a flux-tunable mutual inductance M($\Phi$), with a coupling coefficient $k=\frac{M}{L-M}=\frac{\beta_L\cos\delta_0}{2+\beta_L\cos\delta_0}$, where $\beta_L=2L/L_{J0}<1$, $L_{J0}$ is the junction's Josephson inductance, and $\delta_0$ is the dc phase across the junction determined by $\delta_0+\beta_L\sin\delta_0=2\pi\Phi/\Phi_0$. The effective mutual inductance is positive at $\Phi=0$, negative at $\Phi=\Phi_0/2$, and zero at $\Phi=(\Phi_0/2\pi)\times(\beta_L+\pi/2)$. Figure \ref{fig1}b shows the rf-SQUID effective coupling coefficient $k$ as a function of flux for a device with $\beta_L=0.9$.

Because the mutual inductance associated with the rf-SQUID coupler is tunable through zero, it is useful as a switch element in circuits that rely on this mutual inductance for signal transmission between the input and output ports. While it is tempting to connect the rf-SQUID in-line with a transmission line to implement a microwave switch \cite{Berkley14}, such a device will suffer from high return loss in its `on' state and will in general be narrow-band because the grounded inductors $L$ of the SQUID present the input and output ports with low impedance reactive shunts, and the coupling provided by the SQUID is limited: $k<1/3$ for a mono-stable rf-SQUID at zero flux. We can nevertheless design a microwave switch that does not suffer from these limitations: we observe that the circuit topology of the rf-SQUID coupler, essentially an inductive $\pi$-section, is common in band-pass coupled-resonator filters. We proceed by designing such a filter and embedding the SQUID coupler in place of one of the filter's admittance inverters \cite{MYJ}. The calculated S-parameters of the resulting circuit are shown in Figure \ref{fig1}c.

The design of a band pass filter to embed the SQUID coupler follows the procedure outlined in Ref.\ \onlinecite{MYJ}. The filter's $N$ poles are implemented as parallel $LC$ shunt resonators whose resonant frequency coincides with the filter center frequency $\omega_{res}=\left(L_{res}C_{res}\right)^{-1/2}$. We start with all resonators being identical, and their impedance $Z_{res}=\sqrt{L_{res}/C_{res}}$ will be determined later. The resonators are coupled via admittance inverters (or J-inverters), whose admittance values $\left\{J_{i,j}\right\}$ are calculated from the set of filter coefficients $\left\{g_{i=0\dots N+1}\right\}$ that can be found in design tables, to implement the desired filter characteristics:
$J_{0,1}=\sqrt{\textit{w}/\left(g_0 g_1 Z_{res} Z_{0}\right)}$, $J_{i,i+1}=\textit{w}/\left(Z_{res}\sqrt{g_i g_{i+1}}\right)$, and $J_{N,N+1}=\sqrt{\textit{w}/\left(g_N g_{N+1} Z_{res} Z_0\right)}$, where $\textit{w}$ is the desired fractional bandwidth of the filter, and $Z_0$ is the embedding impedance, usually 50 $\Omega$.

We implement all but one of the admittance inverters as capacitive $\pi$-sections with $C_{i,j}=J_{i,j}/\omega_{res}$. The remaining inverter $J_{k,k+1}$ is implemented with an inductive $\pi$-section and accommodates the SQUID coupler. This sets the Josephson inductance of the coupler junction $L_{J0}=1/\omega_{res}J_{k,k+1}$. The negative shunt capacitances and inductances associated with the inverters in the interior sections of the filter are in parallel with, and can be absorbed by, those of the shunt resonators. The negative capacitances of the first and last inverters can be replaced by a series capacitor near the termination and a negative shunt capacitor in parallel with the first and last resonators, so that the first (last) series capacitance becomes $C_s=J/\omega_{res}\sqrt{1-(Z_0 J)^2}$, and the associated (negative) shunt capacitance becomes $C_e=J\sqrt{1-(Z_0 J)^2}/\omega_{res}$, where $J$ is the admittance of the first (last) inverter \cite{MYJ}.

The characteristic impedance of the shunt resonators is constrained by two conditions: first, that the SQUID coupler have $\beta_L<1$, favoring low impedance resonators. The second constraint, which favors high impedance resonators, follows from the requirement that the first and last series capacitances remain finite, i.e. $Z_0\times\max\left\{J_{0,1},J_{N,N+1}\right\}<1$. Both constraints can be satisfied simultaneously in most filter topologies.

Our microwave switch design, shown in Figure \ref{fig1}a, is based on a two pole maximally flat filter prototype having 27\% fractional bandwidth with a center frequency of 10 GHz. The resonator impedances were set to 11.6 $\Omega$ and their capacitances, after absorbing the negative shunt capacitances of the inverters, are $C_{r1}=C_{r2}=1.19$ pF. The resonator inductances, after absorbing the negative shunt inductance of the inverter, are $L_1=L_2=229$ pH. The junction critical current is $I_c=0.88~\mu$A, giving $L_{J0}=374$ pH. The junction is connected to $L_1$ and $L_2$ at a tap 169 pH above ground, keeping the rf-SQUID mono-stable with $\beta_L=0.9$. The series capacitors are $C_{c1}=C_{c2}=0.659$ pF.

We have measured two devices with nominally identical parameters, but with different geometries for the rf-SQUID inductors $L_1$ and $L_2$. In Device 1, a CAD drawing of which is shown in Figure \ref{fig1}d, the inductors were implemented as multilayer gradiometric solenoids, and in Device 2 the inductors were implemented as planar spirals. The couplings to the flux bias line were designed to be 8 pH and 2.8 pH and were measured to be 6.6 pH and 2.9 pH for Devices 1 and 2, respectively. The devices were fabricated at Northrop Grumman with Al/AlO$_x$/Al Josephson junctions and Nb wiring, and were measured in a dilution refrigerator at a base temperature of approximately 20 mK. The two devices showed similar results.

\begin{figure}
\includegraphics[width=\columnwidth]{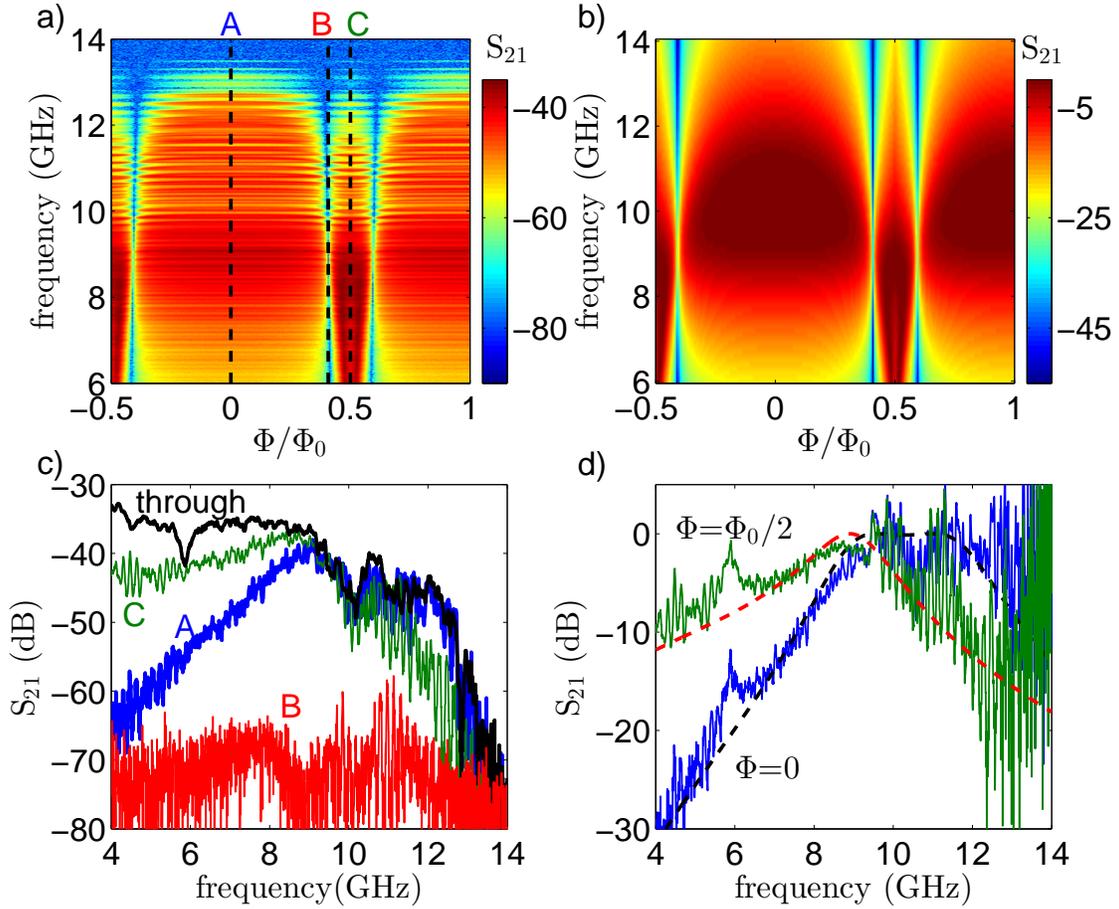}
\caption{\label{fig2}a) Measured transmission S$_{21}$ (in dB) vs applied flux, $P_{in}=-120$ dBm, for Device 1, b) simulated S$_{21}$ vs flux, c) S$_{21}$ vs frequency at $\Phi/\Phi_0 =0$ (A), 0.407 (B), and 0.5 (C), as well as a through calibration trace, d) S$_{21}$ normalized to the through data, overlaid with simulated data (dashed).}
\end{figure}

Figure \ref{fig2}a shows the transmission through Device 1, S$_{21}(\omega)$, as a function of applied flux. S-parameters are referred to the network analyzer ports, and include approximately 100 dB attenuation to the input of the device, and approximately 60 dB of gain following the device's output port. The input power to the device was nominally $-120$ dBm. The data is periodic in flux and shows two regions of high transmission separated by transmission nulls. We associate the wide transmission region with the positive coupling lobe of the rf-SQUID around integer flux quanta, and the narrow region with the negative coupling lobe (see Figure \ref{fig1}b) around half-integer flux quanta, where the signal is transmitted with a 180 degree phase shift. From the positions of the transmission nulls, which occur at $\Phi/\Phi_0=0.407$ and 0.389 in devices 1 and 2 respectively, we can extract the rf-SQUID $\beta_L=0.99$ in Device 1 and $\beta_L=0.85$ in Device 2. Panel b) in the figure shows an S-parameter simulation of the circuit using the designed values of passive components and junction inductances calculated from the experimentally determined $\beta_L$.

Figure \ref{fig2}c shows S$_{21}(\omega)$ line-cuts through the dataset of panel a) at different flux biases as indicated by the dashed lines. The traces are compared to a through S$_{21}$ trace obtained with a short coaxial line replacing the device on the same measurement chain. Trace A in Figure \ref{fig2}c represents the `on' position of the Josephson microwave switch at $\Phi/\Phi_0=0$. We see that the transmission in this position essentially coincides with that of the through trace over the designed pass-band of the embedding filter, from 9-11 GHz. Trace B in the figure represent the switch's `off' position at $\Phi/\Phi_0=0.407$, showing broad-band suppression of the transmission through the device, with an on/off ratio greater than 20 dB over the device's entire pass-band. In this position the signal energy is reflected to the input port. Trace C shows the response of the switch when biased in the negative coupling regime near $\Phi/\Phi_0=0.5$. The maximum transmission is still close to unity, however, the designed filter characteristics are not preserved because of the negative inductance of the admittance inverter containing the junction. Panel d) in the figure compares the `on' and `negative' S$_{21}$ traces, normalized to the through data, with simulated S-parameters. Good agreement between experiment and simulation validates our design concept and methodology for this type of device. The continuous variation of the transmission with SQUID flux bias from near unity, through zero, and to near unity but with 180 degree phase shift, makes this device useful not only as a 2-state switch but also as a scalar modulator.  

\begin{figure}
\includegraphics[width=3.2in]{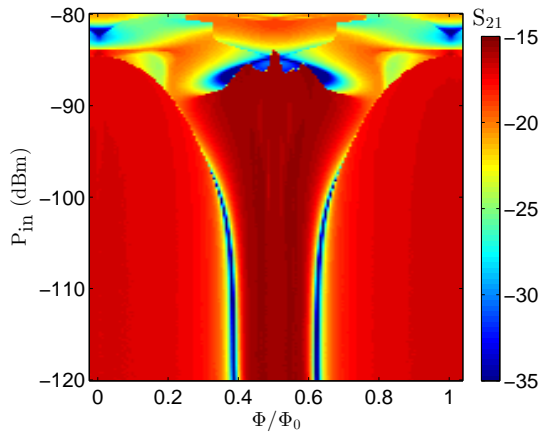}
\caption{\label{fig3}Power-swept S$_{21}$ at 9 GHz vs flux for Device 2. S$_{21}$=-15 dB represents full transmission.}
\end{figure}

We proceed to characterize the large-signal response of the device. Figure \ref{fig3} shows S$_{21}$ CW-frequency power sweeps of Device 2 at 9 GHz as a function of applied flux. The quoted input power is the nominal power referred to the device input port. S$_{21}=-15$ dB in the figure corresponds to unity transmission through the device. We see that for input powers up to approximately $P_{in}=-105$ dBm the device still operates in its small-signal, linear regime. When the power is increased further the nonlinearity of the junction becomes important, which is evident by the shift of the transmission null towards integer flux. At powers greater than $-100$ dBm the transmission null becomes progressively more transparent, until it eventually disappears, as the rf-SQUID appears to transition discontinuously from positive to negative coupling. At $P_{in}>-90$ dBm the device becomes globally unstable, first in the negative coupling lobe, and later in the positive lobe as well.

While a shift of the transmission null towards integer flux is not surprising and is consistent with an increase of the nonlinear junction inductance with drive amplitude \cite{Vijay09}, we do not have at the time of this writing a quantitative model to understand the power dependence of Figure \ref{fig3}. Simulations in WRSpice \cite{Whiteley91} agree with the data on the overall trend, but differ significantly in the range of applied powers over which the nonlinear behavior is observed.

Figure \ref{fig3} demonstrates that the device is useful as a microwave switch for input powers up to approximately -100 dBm. This is sufficient to accommodate signal levels associated with qubit readout, and to drive Rabi oscillations in a typical transmon qubit through a 1 fF gate capacitor with $\Omega_{Rabi}>50$ MHz \cite{Koch07}. The third order intercept power for Device 1, with 100 kHz offset signals at 8.5 GHz and $\Phi/\Phi_0=0.5$, was measured to be $-72.5$ dBm referred to the input. Higher power handling capabilities may be achieved if the single junction is replaced with a series array of N junctions, each having N times the critical current of the single junction.

\begin{figure}
\includegraphics[width=3.0in]{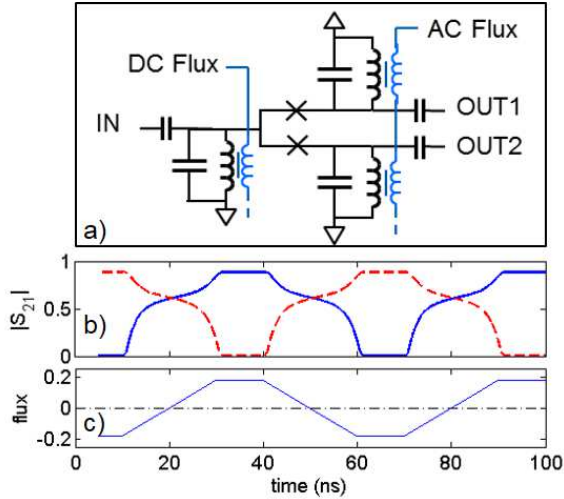}
\caption{\label{fig4}a) Schematic of an SPDT switch, b) simulated transmission $\left|S_{21}\right|$ from the input port to ports OUT1 (dashed) and OUT2 (solid) of the SPDT switch in response to the ac flux control signal in c), shown in units of $\Phi_0$, with the dc flux bias held at 0.18 $\Phi_0$.}
\end{figure}

In Figure \ref{fig4} we show how the same design principles that we have demonstrated here can be applied to a single-pole double-throw (SPDT) switching device, which can be used to form larger scale microwave binary tree multiplexers, and offers better routing flexibility in comparison to the simple SPST switch. The device is still a 2-pole band pass filter as in Figure \ref{fig1}a, however, here we duplicated the second pole resonator to provide an additional output, and connected this additional stage via a second junction to the input resonator. The two junctions, together with the shunt inductors of the three resonators form two rf-SQUID loops sharing a common inductance. The flux enclosed in each of the loops can be controlled independently with two flux lines as shown in panel a), but it is convenient to set one of the fluxes at a dc value and use the other flux line to toggle the switch. In this way the total flux in one of the loops can be set to zero, while the flux in the other loop can be set to the coupler's null position. We performed transient anlaysis of the SPDT switch in WRSpice with identical component values to those used in the SPST switch reported above, and with a -120 dBm, 9 GHz input signal. The time domain voltage at the output ports was numerically demodulated to obtain the signal magnitude intergrated over 10 cycles of the carrier. Figure \ref{fig4}b shows the magnitude of the simulated transmission from the input port to either of the output ports in response to the 0.18 $\Phi_0$-per-junction ac flux signal shown in panel c), and with the dc flux bias held at 0.18 $\Phi_0$. We see that the input signal is routed to either one of the two outputs while the other, unselected output is isolated. The input port remains matched when either of the outputs is engaged. The switching speed is not limited to the 20 ns simulated here, but in practice it is desirable to keep the frequency of the ac control flux below the cutoff frequency of the signal path to avoid excessive IF-RF leakage. 

To summarize, we have demonstrated a flux-controlled microwave switch based on a Josephson junction coupler embedded in a band-pass filter, with low insertion loss, over 2 GHz instantaneous bandwidth, and over 20 dB on/off ratio. We observe good agreement with simulations in the device's linear regime for input powers up to approximately $-100$ dBm. We demonstrated by simulation an extension of the basic device to an SPDT switch configuration that enables flexible routing and integration into switch matrices. We presented a design outline that makes it possible to implement similar devices using different filter prototypes or different number of poles, with fractional bandwidths up to 40\%. These devices represent a new family of non-dissipative Josephson junction circuits for integrated routing and modulation of microwaves for qubit control and readout.



%
%

%


\bibliography{switch_APL}

\end{document}